\documentclass[tc, 2-Column Paper]{copernicus}

\begin{document}

\title{Radiative transfer model for contaminated slabs : experimental validations}

% \Author[affil]{given_name}{surname}

\Author[1,2]{Andrieu}{Fran\c{c}ois}
\Author[1,2]{Schmidt}{Fr\'{e}d\'{e}ric}
\Author[3]{Schmitt}{Bernard}
\Author[3]{Dout\'{e}}{Sylvain}
\Author[3]{Brissaud}{Olivier}

\affil[1]{Universit\'{e} Paris-Sud, Laboratoire GEOPS, UMR8148, Orsay F-91405, France}
\affil[2]{CNRS, Orsay F-91405, France}
\affil[3]{Institut de Plan\'{e}tologie et d'Astrophysique de Grenoble, Grenoble 38041, France}

%% The [] brackets identify the author with the corresponding affiliation. 1, 2, 3, etc. should be inserted.

\runningtitle{Radiative transfer model for contaminated slabs : experimental validations}

\runningauthor{Andrieu}

\correspondence{ANDRIEU F. (francois.andrieu@u-psud.fr)}

\received{}
\pubdiscuss{} %% only important for two-stage journals
\revised{}
\accepted{}
\published{}

%% These dates will be inserted by Copernicus Publications during the typesetting process.

\firstpage{1}

\maketitle

\begin{abstract}
This article presents a set of spectro-goniometric measurements of
different water ice samples and the comparison with an approximated
radiative transfer model. The experiments were done using the spectro-radiogoniometer
described in \cite{Brissaud2004}. The radiative transfer
model assumes an isotropization of the flux after the second interface
and is fully described in \cite{Andrieu2015}. 

Two kind of experiments were conducted. First, the specular spot was closely investigated, at high angular resolution, at the wavelength of $1.5\,\mbox{\ensuremath{\mu}m}$ , where ice behaves as a very absorbing media. Second, the bidirectional reflectance was sampled at various geometries, including low phase angles on 61 wavelengths ranging from $0.8\,\mbox{\ensuremath{\mu}m}$
  to $2.0\,\mbox{\ensuremath{\mu}m}$
 . 

In order to validate the model, we made a qualitative test to demonstrate the relative isotropization of the flux. We also conducted quantitative assessments by using a bayesian inversion method in order to estimate the parameters (e.g. sample thickness, surface roughness) from the radiative measurements only. A simple comparison between the retrieved parameters and the direct independent measurements allowed us to validate the model. 

We developed an innovative bayesian inversion approach to quantitatively estimate the uncertainties on the parameters avoiding the usual slow Monte Carlo approach. First we built lookup tables, and then searched the best fits and calculated a posteriori density probability functions. The results show that the model is able to reproduce the spectral behavior of water ice slabs, as well as the specular spot. In addition, the different parameters of the model are compatible with independent measurements. 
\end{abstract}

\introduction  %% \introduction[modified heading if necessary]

Various species of ices are present throughout the Solar System. From
water ice and snow on Earth to nitrogen ice on Triton \citep{Zent89},
not forgetting carbon dioxide ice on Mars \citep{Leighton1966}. Ice an snow covered areas have a strong impact on planetary climate dynamics, as they can lead to significant regional scale albedo changes at the surface and surface/atmosphere volatiles  interactions. The physical properties of the cover have also an impact on the energy balance : for example the albedo depend on the grain size of the snow \citep{dozier2009, negi2011}, on the roughness of the interface \citep{lhermitte2014} on the presence or not and the physical properties of impurities \citep{dumont2014}, or on the specific surface area \citep{Picard2009, mary2013}. The study and monitoring of theses parameters is a key to constrain the energy balance of a planet.

Radiative transfer models have proven essential to retrieve such properties and their evolution at a large scale, and different families exist. Ray tracing algorithms, such as those described in \cite{Picard2009} for snow or \cite{Pilorget2013} for compact polycrystalline ice simulate the complex path of millions of rays into the surface. They provide very accurate simulations, but have the weakness of being time consuming. Analytical solutions of the radiative transfer in homogeneous granular media have been developed for example by \cite{Shkuratov1999} and \cite{Hapke1981}. They are fast, and when the surface cannot be described as homogeneous, they must be combined with another family of techniques such as discrete ordinate methods like DISORT \citep{Stamnes1988}.
These methods have been widely studied on Earth snow \citep{carmagnola2013, dozier2009, dumont2010, painter2004} and other planetary cryospheres \citep{Appere2011, Eluszkiewicz2003}, modeling a granular surface. Compact polycrystalline ices have however been recognized to exist
on several objects : CO$_{2}$ on Mars \citep{Kieffer2001,Eluszkiewicz2005},
N$_{2}$ on Triton and Pluto, \citep{Zent89,Eluszkiewicz2003} and probably
SO$_{2}$ on Io \citep{Eluszkiewicz2003}, owing to the very long light
path-lengths measured, over several decimeters. Radiative transfer in compact slabs is different from in granular media. We developed an approximated
model \citep{Andrieu2015}  designed to study contaminated ice slabs,
with a fast numerical implementation, that has already been
numerically validated. The main objective of the model is the analysis
of massive spectro-imaging planetary data of these surfaces. For this
purpose, it is semi analytic and fast implemented. It is designed to retrieve the variations  of thickness and impurity content of compact polycrystalline planetary ices.

In the present article, we will test the accuracy of this approximated
model on laboratory spectroscopic measurements of pure water ice Bidirectional Reflectance
Distribution Function (BRDF). The goal is to propose an inversion
framework to retrieve surface properties, including uncertainties
in order to demonstrate the validity of the approach. In order to
speed up the inversion, we based the algorithm on look-up tables that minimize the computation time of the direct model. This strategy
will be very useful to analyze real hyperspectral images. The thickness
of ice estimated from the inversion is validated in comparison to
real direct measurements. Also the specular lobe is adjusted to demonstrate
that the model is able to reasonably fit the data with a coherent
roughness value.

\begin{figure}[t]
\includegraphics[width=8.3cm]{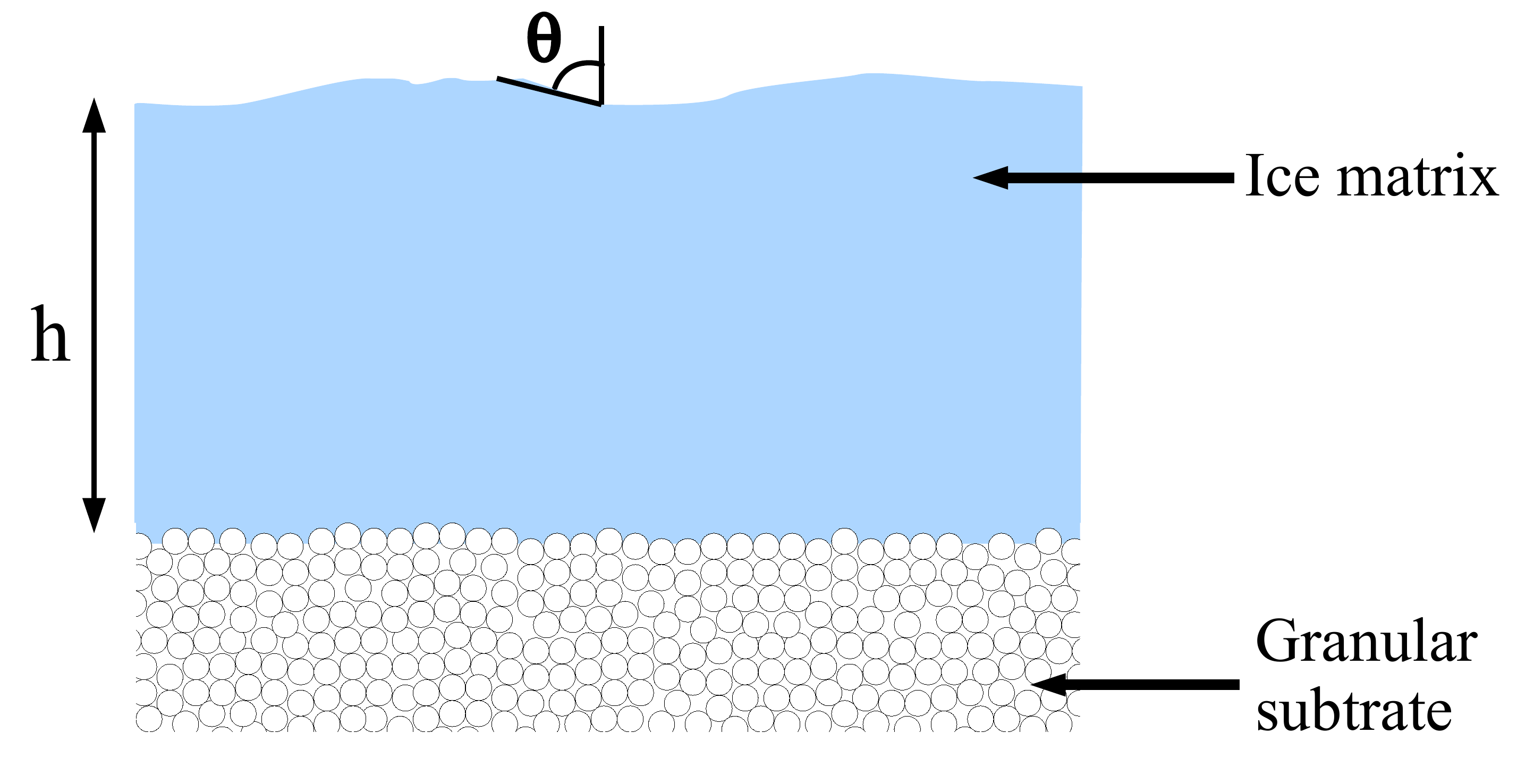}
\caption{Scheme of the surface representation in
the radiative transfer model applied on the laboratory measurements.
$h$ represents the slab thickness. $\bar{\theta}$ represent the
mean slope to describe the surface roughness.}
\label{fig:Scheme-of-the}
\end{figure}

\section{Description of the model}
\label{sec:Descrition-of-the}
The model,  \cite{Andrieu2015},  is inspired from an existing one described in  \cite{Hapke1981} and \cite{Doute1998}, that simulates the bidirectional reflectance of stratified granular
media. It has been adapted to compact slab, contaminated with pseudo-spherical
inclusions, and a rough top interface. In the context of this work,
we suppose a layer of pure slab ice, overlaying an optically thick
layer of granular ice, as described on figure \ref{fig:Scheme-of-the}.
The roughness of the first interface is described using the probability
density function of orientations of slopes defined in \cite{Hapke1984}.
This distribution of orientations is fully described by a mean slope
parameter $\bar{\theta}$. The ice matrix is described using its optical
constants and it thickness. Within the slab, the model can also incorporate
inclusions, supposed to be close to spherical and homogeneously distributed
inside the matrix. They are described by their optical constants,
their volumetric proportions and their characteristic grain-sizes.
There can be several different types of inclusions. Each type can
be of any material, except the one constituting the matrix : it can
be any other kind of ice, mineral, or even bubbles. 
\begin{figure}[t]
\includegraphics[width=8.3cm]{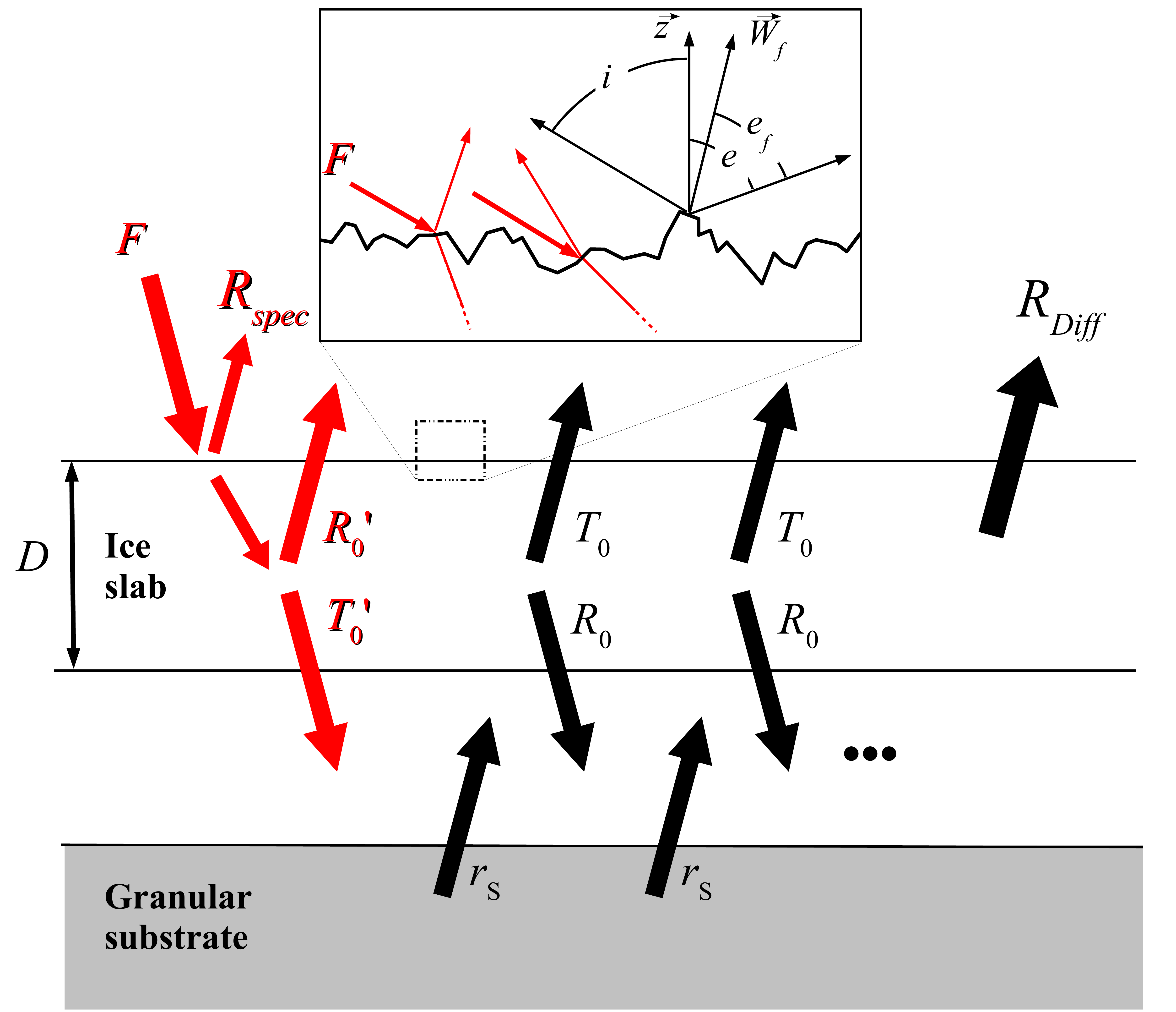}
\caption{Illustration of the radiative
transfer in the surface. Anisotropic transits are represented in red.
$\vec{F}$ is the incident radiation flux, $R_{spec}$ and $R_{Diff}$
are respectively the specular and diffuse contributions to the reflectance
of the surface. $r_{s}$ is the lambertian reflectance of the granular
substrate, and $R_{0}$ and $T_{0}$ are total reflexion and transmission
factors of the slab layer. A prime indicates an anisotropic transit.
The reflection and transmission factors are different in isotropic
or anisotropic conditions. The granular and slab layers are artificially
separated in this figure to help the understanding of the coupling
between the two layers. On the top : illustration of the reflections
and transmission at the first interface, used in the calculations
of $R_{spec}$ and the determination of the amount of energy injected
in the surface. $\vec{z}$ is the normal to the surface, $\vec{W}_{f}$
the local normal to a facet, $i$ and $e$ are respectively the incidence
and emergence angle, and $e_{f}$ is the local emergence angle for
a facet. Each different orientation of a facet will lead to a different
transit length in the slab. A more detailed description can be found in \cite{Andrieu2015}.}
\label{fig:Illustration-of-radiative}
\end{figure}

Figure \ref{fig:Illustration-of-radiative} illustrates the general
principle of the model. The simulated bidirectional reflectance results
from two separate contributions: specular and diffuse. The specular
contribution of a measurement is estimated from the roughness parameter,
the optical constants of the matrix, and the apertures of the light
source and the detector. The surface is considered as constituted
of many unresolved facets, which orientations follow the defined probability
density function. The specular reflectance is obtained integrating
every reflection on the different facets. The total reflection coefficient
at the first rough interface is obtained by integrating specular contributions
in every emergent direction, at a given incidence. This gives the
total amount of energy transmitted into the system constituted of
the contaminated slab and the substrate. The diffuse contribution
is then estimated solving the radiative transfer equation inside this
system under various hypothesis. It is considered that : (i) the first
transit through the slab is anisotropic due to the collimated radiation
from the source, and that there is an isotropization at the second
rough interface (\textit{i.e. }when the radiation reach the semi-infinite
substrate). For the refraction and the internal reflection, every
following transit is considered isotropic. (ii) the geometrical optics
are valid. This means that the size of the inclusions and the thickness
of the slab layer must be larger than the considered wavelength. (iii)
the inclusions inside the matrix are close to spherical and homogeneously
distributed. The reflexion and transmission factors of the layers
are obtained using an analytical estimation of Fresnel coefficients described in
\cite{Chandrasekhar1960,Doute1998}, and a simple statistical approach, detailed in 
\cite{Andrieu2015}. The contribution of the semi infinite substrate
is described by its single scattering albedo. Finally, as the slab
layer is under a collimated radiation from the light source, and under
a diffuse radiation from the granular substrate, the resulting total
bidirectional reflectance is computed using adding doubling formulas.

\section{Data}
\label{sec:Data}

\subsection{Spectro-radiogoniometer}
\label{sub:Spectro-radiogoniometer}

\begin{figure}[t]
\center{\includegraphics[height=7.3cm]{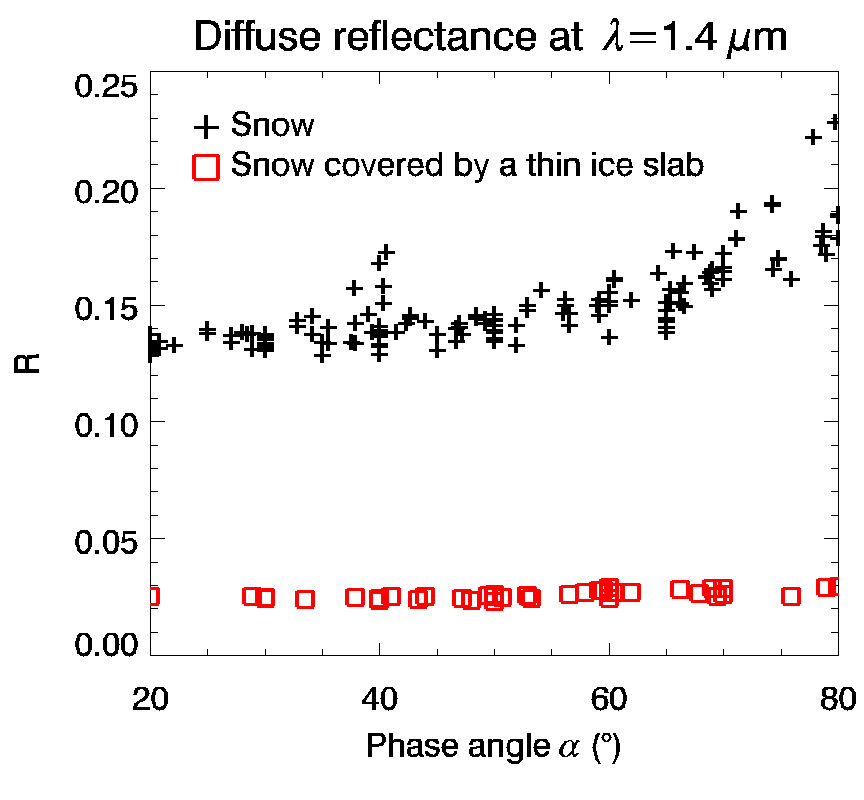}}
\newline
\centering{(a)}
\center{\includegraphics[height=7.3cm]{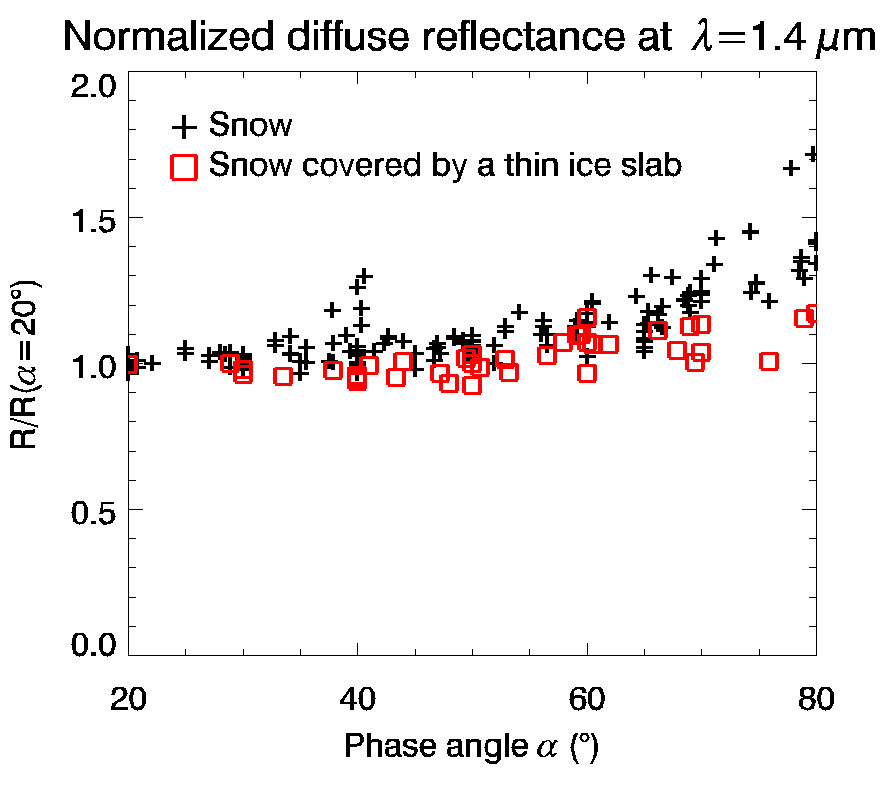}}
\centering{(b)}
\caption{(a) Reflectance factor at a wavelength
$\lambda=1.4\,\mbox{\ensuremath{\mu}m}$versus phase angle for snow only (black crosses)
and the same snow, but covered with a $1.42\pm0.27\,\mbox{mm}$ water
ice slab (red squares). The thin layer of slab ice not only lower
the level of reflectance as expected, but also seems to isotropize
the reflected radiation. It is clearer on plot (b), that represents
the same data, but normalized by the value at a phase angle $\alpha=20\text{\textdegree}$.
This data shows that even a very thin layer of ice has a strong effect
on the directivity of the surface. This justify the approximation
of isotropization at the second interface supposed by the model, and
the description of the substrate using only its single scattering
albedo.}
\label{fig:Reflectance-snow-vs-slab}
\end{figure}

The bidirectional reflectance spectra were measured using the spectrogonio-radiometer
from IPAG fully described in \cite{Brissaud2004}. We collected spectra in the near infrared
at incidences ranging from $40\text{\textdegree}$ to $60\text{\textdegree}$,
emergence angles from $0\text{\textdegree}$ to $50\text{\textdegree}$,
and azimuth angles from $0\text{\textdegree}$ to $180\text{\textdegree}$.
The sample is illuminated with a large monochromatic beam (divergence
$<1\text{\textdegree}$) an the near infrared spectrum covering the
range from $0.800\,\mbox{\ensuremath{\mu}m}$ to $4800\,\mbox{\ensuremath{\mu}m}$ is measured by an InSb
photovoltaic detector. This detector has a nominal aperture of $4.2\text{\textdegree}$,
that result in a field of view on the sample of approximately $20\, mm$
in diameter. The minimum angular sampling of illumination and observation
directions is $0.1\text{\textdegree}$, with a reproducibility of
$0.002\text{\textdegree}$. In order to avoid azimuthal anisotropy,
the sample is rotated during the acquisition. The sample rotation
axis may be very slightly misadjusted, resulting on a notable angular drift on the emergence measured up to $1\text{\textdegree}$.

\subsection{Ice BRDF measurements}
\label{sub:Ice-slabs-measurements}
The ice samples were obtained by sawing artificial columnar water
ice in sections. These sections were put on top of an optically thick
layer of compacted snow, that was collected in Arselle, in the French
Alps. The spectral measurements were conducted in a cold chamber at
263K. Still, the ice and the snow were unstable in the measurement's
environment, due to the dryness of the chamber's atmosphere. The grain-size
of the snow showed an evolution and the thickness of a given slab
showed a decrease of $0.343\,\mbox{mm/day}$. Each sample needs a
acquisition time of $10\,\mbox{hours}$. For each measurement, the
ice slab was sliced, and its thickness was measured in five different
locations. It was then set on top of the snow sample, and this system
was put into the spectro-goniometer, and put into rotation for the
measurement. As the surface is not perfectly plane, the measured thickness
is not constant. This results in an $2\sigma$ standard-deviation
in the measurement of the thickness than ranges from $0.54\,\mbox{mm}$
to $2.7\,\mbox{mm}$ in our study, depending on the sample.

\paragraph{Specular contribution\protect \\
}

The specular contribution was measured on a $12.51\,\mbox{mm}$ thick
slab sample on top of Arselle snow. This sample is the same sample
that sample 3 described in the next paragraph. The illumination was
at an incidence angle of $50\text{\textdegree}$, and 63 different
emergent geometries were sampled, ranging from $45\text{\textdegree}$
to $55\text{\textdegree}$ in emergence, and from $170\text{\textdegree}$
to $180\text{\textdegree}$ in azimuth. A measure at the wavelength
of $1.5\,\mbox{\ensuremath{\mu}m}$ is represented on figure \ref{fig:Specular2D} (a).
The sampling is $1\text{\textdegree}$ in emergence and azimuth within
$47\text{\textdegree}$ and $53\text{\textdegree}$ in emergence and
$175\text{\textdegree}$ and $180\text{\textdegree}$ in azimuth.

\paragraph{Diffuse reflectance spectra\protect \\
}

The diffuse contribution was measured on three samples of different
slab thickness. The three thicknesses were measured on different locations
of the samples with a caliper before the spectrogonio measurement,
resulting in $h_{1}=1.42\pm0.47\,\mbox{mm}$, $h_{2}=7.45\pm0.84\,\mbox{mm}$,
$h_{3}=12.51\pm2.7\,\mbox{mm}$, respectively for samples 1, 2 and
3, with errors at $2\sigma$. $61$ wavelengths were sampled ranging
from $0.8\,\mbox{\ensuremath{\mu}m}$ to $2.0\,\mbox{\ensuremath{\mu}m}$. Spectra were collected on 39 different points of the BRDF, respectively for the incidence, emergence
and azimuth angles : $\left[40\text{\textdegree},50\text{\textdegree},60\text{\textdegree}\right]$,
$\left[0\text{\textdegree},10\text{\textdegree},20\text{\textdegree}\right]$
and $\left[0\text{\textdegree},45\text{\textdegree},90\text{\textdegree},140\text{\textdegree},160\text{\textdegree},180\text{\textdegree}\right]$.
This set of angles results in only 39 different geometries because
the azimutal angle is not defined for a nadir emergence.

\paragraph{Snow diffuse reflectance spectra\protect \\
}

Diffuse reflectance spectra of natural snow only were also measured.
The objective was to estimate the effect of a slab layer on the BRDF.
Figure \ref{fig:Reflectance-snow-vs-slab} show on the same plots
the reflectance factor versus phase angle of the snow and the snow
covered with a $1.42\,\mbox{mm}$ thick ice slab (sample 1). It illustrates
the two most notable effects of a thin layer of slab ice on top of
an optically thick layer of snow. The most intuitive effect is to
lower the level of reflectance : it is due to absorption during the
long optical path lengths in the compact ice matrix. The second is
that the radiation is more lambertian than the one of only snow. This
data gives credit to the first hypothesis of isotropization of the
radiation formulated in the model (see Section \ref{sec:Descrition-of-the}).
The description of the bottom granular layer as isotropic, defined
only its single scattering albedo, may be considered brutal, but this
dataset shows that a thin coverage of slab ice even on a very directive
material such as snow, is enough to strongly flatten the BRDF.

\section{Method\label{sec:Inversion-method}}

We designed an inversion method aiming at massive data analysis. This
method consists in two steps : first the generation of a synthetic
database that is representative of the variability of the model, and
then comparison with actual data. To generate the synthetic database,
we used optical constants for water ice at $270\,\mbox{K}$. The $7\,\mbox{K}$
difference between the actual temperature of the room and the temperature
assumed for the optical constants has negligible effect. We combined
the datasets of \cite{Warren2008} and \cite{Schmitt1998}
making the junction at $1\,\mbox{\ensuremath{\mu}m}$, the former
set for the shorter wavelengths, and the latter for the wavelengths
larger than $1\,\mbox{\ensuremath{\mu}m}$. 

In order to validate the model on the specular reflection from the
slab, we chose to use the reflectance at $1.5\,\mbox{\text{\ensuremath{\mu}}m}$,
where the ice is very absorbing. Figures \ref{fig:Measured-and-simulated1},
and \ref{fig:Specular2D} clearly
demonstrate that there is negligible diffuse contribution in geometry
outside the specular lobe from the sample with $12.51\,\mbox{mm}$
thick pure slab. Thus, the roughness parameter $\bar{\theta}$ is
the only one impacting the reflectance in the model. We chose to inverse
this parameter first, and validate the specular contribution.

We will then focus on the validation in the spectral domain, for the
diffuse contribution. We will use the estimation of the roughness
parameter $\bar{\theta}$ obtained earlier and the spectral data in
order to estimate the slab thickness and the grain-size of the snow
substrate. To do this, we assumed that the roughness was not changing
significantly enough to have a notable impact on diffuse reflectance
from one sample to another. This assumption is justified by the fact
that the different columnar ice samples were made the same way, as
flat as possible and the low value of $\bar{\theta}$ retrieved as
discussed in the next section. It is confirmed by the results of section
\ref{sub:Specular}, that suggest a very low roughness, as expected.
Such low roughness parameter have negligible influence on the amount
of energy injected into the surface.

\subsection{Inversion strategy}

The inversion consists in estimating the model parameter $m$ from
the model $F(m)$, that are close to the data $d$. \cite{Tarantola1982} showed that it can be mathematically solved by considering each element as a Probability Density Function
(PDF) . In non-linear direct problem, the solution
may not be analytically approached. Nevertheless, it is possible to
sample the solution PDF with a Monte Carlo approach as shown in  \cite{Mosegaard1995},
but this solution is very time consuming. 

The actual observation is considered as prior information on data
$\rho_{D}(d)$ in the observation space $D$. It is assumed to be
a $N$-dimension gaussian PDF $\mathcal{G}(d_{mes},\overline{\overline{C}})$,
with mean $d_{mes}$ and covariance matrix $\overline{\overline{C}}$.
The values $r_{i}$ are the observations for each element (angular
or spectral as described later). The covariance matrix $\overline{\overline{C}}$
is assumed here to be diagonal since measurements at a given geometry/wavelength
are independent of the other measurements. The diagonal elements $C_{ii}$
are $\sigma_{1}^{2},\ldots,\sigma_{N}^{2}$, with $\sigma_{i}$ being
the standard deviation. The prior information on model parameters
$\rho_{M}(m)$ in the parameter space $M$ is independent to the data
and corresponds to the state of null information if no information
is available on the parameters. We consider an uniform PDF in their
definition space $M$. The state of null information $\mu_{D}(d)$
represents the case when no information is available. It is trivial
in our case and represent the uniform PDF in the parameters space
$M$. The posterior PDF in the model space $\sigma_{M}(m)$ as defined
by \cite{Tarantola1982} is : 
\begin{equation}
\sigma_{M}(m)=k\rho_{M}(m)L(m)
\end{equation}
where $k$ is a constant and $L(m)$ is the likelihood function
\begin{equation}
L(m)=\int_{D}\mbox{d}d\,\frac{\rho_{D}(d)\,\theta(d\mid m)}{\mu_{D}(d)}
\end{equation}
where $\theta(d\mid m)$ is the theoretical relationship of the PDF
for $d$ given $m$. We do not consider errors on the model itself,
so $\theta(d\mid m)=\delta(F(m))$ also noted $d_{sim}$ for simulated
data. So the likelihood is simplified into :

\begin{equation}
L(m)=\mathcal{G}(F(m)-d_{mes},\overline{\overline{C}})
\end{equation}
and in the case of an uniform prior information $\rho_{M}(m)$, the
posterior PDF is:

\begin{equation}
\sigma_{M}(m)=kL(m)
\end{equation}
this expression is explicitly :
\begingroup\makeatletter\def\f@size{9}\check@mathfonts
\def\maketag@@@#1{\hbox{\m@th\large\normalfont#1}}
\begin{equation}
\sigma_{M}(m)=k.\exp\left(-\frac{1}{2}\times{}^{t}\left(F(m)-d_{mes}\right)\overline{\overline{C}}^{-1}\left(F(m)-d_{mes}\right)\right)\label{eq:Ltheo}
\end{equation}
\endgroup
The factor $k$ is adjusted to normalize the PDF. The mean value of
the estimated parameter can be computed by :

\begin{equation}
\left\langle m\right\rangle =\int_{M}m.\sigma_{M}(m)\mbox{d}m\label{eq:mean}
\end{equation}
and the standard deviation:

\begin{equation}
\sigma_{\left\langle m\right\rangle }=\int_{M}\left(m-\bar{m}\right)^{2}.\sigma_{M}(m)\mbox{d}m\label{eq:standardev}
\end{equation}
In order to speed up the inversion strategy but keep the advantage
of the Bayesian approach, we choose to sample the parameter space
$M$ with regular and reasonably fine steps, noted $i$. The likelihood
for each element is:
\begingroup\makeatletter\def\f@size{9}\check@mathfonts
\def\maketag@@@#1{\hbox{\m@th\large\normalfont#1}}
\begin{equation}
L(i)=\exp\left(-\frac{1}{2}\times{}^{t}\left(d_{sim}(i)-d_{mes}\right)\overline{\overline{C}}^{-1}\left(d_{sim}(i)-d_{mes}\right)\right)\label{eq:L}
\end{equation}
\endgroup
The derivation of posterior PDF with such formalism for specular lobe
inversion and for spectral inversion are explained in the next sections.

\subsection{Specular lobe\label{sub:Specular}}

To study the specular spot, we have to consider the whole angular
sampling of the spot as single data measurement. Similar to the ``pixel''
(contraction of \emph{picture} \emph{element}), we choose to define
the ``angel'' (contraction of \emph{angular} \emph{element}), as
a single element in a gridded angular domain. Interestingly, angel
also refer to a supernatural being represented in various forms of
glowing light. A single angel measurement could not well constrain
the model, even at different wavelengths. Instead a full sampling
around the specular lobe should be enough, even at one single wavelength.
We chose a wavelength where the diffuse contribution was negligible
in order to simplify the inversion strategy. We first generated a
synthetic database (look-up table), using the direct radiative transfer
model. We simulated spectra in the same geometrical conditions, for
a $12.5\,\mbox{mm}$ thick ice layer over a granular ice substrate
constituted of $1000\,\mbox{\ensuremath{\mu}m}$ wide grains. These
two last parameters are not important since the absorption is so high
in ice, such the main contribution is from the specular reflection,
and the diffuse contribution is negligible.

The sampling of the parameters space, that is the lookup table, must
represent correctly every possible variability. For this study, we
sampled the roughness parameter from $0.1\text{\textdegree}$ to $5\text{\textdegree}$
with a constant step $\mbox{d}\bar{\theta}=0.01\text{\textdegree}$.
We use a likelihood function $L$ defined in Eq. \ref{eq:L}, where
$d_{sim}$ and $d_{mes}$ are $n_{geom}$-elements vectors, with $n_{geom}$
the number of angel (63 in that study). They represent respectively
the simulated and measured reflectance at a given wavelength in every
geometry. $\overline{\overline{C}}$ is
a $n_{geom}\times n_{geom}$ matrix. It represents the uncertainties
on the data. In this case, we considered each wavelengths independently,
thus generating a diagonal matrix, containing the level of errors
given by the technical data of the instrument given by \cite{Brissaud2004}.
The roughness parameter $\bar{\theta}$ returned by the inversion
will be described by its normalized PDF: 
\begin{equation}
\mathcal{P}\left\{ \bar{\theta}(i)\right\} =\frac{L\left(i\right)\mbox{d}\bar{\theta}}{\sum_{j}L\left(j\right)\mbox{d}\bar{\theta}}=\frac{L\left(i\right)}{\sum_{j}L\left(j\right)}
\end{equation}
The best match, is the value $\bar{\theta}(i)$ with the highest probability.
The full PDF can be estimated by its mean : 
\begin{equation}
\left\langle \bar{\theta}\right\rangle =\frac{\sum_{i}\bar{\theta}\left(i\right)L\left(i\right)}{\sum_{i}L\left(i\right)}
\end{equation}
and associated standard deviation :
\begin{equation}
\sigma_{\left\langle \bar{\theta}\right\rangle }=\sqrt{\frac{\sum_{i}\left(\bar{\theta}\left(i\right)-\left\langle \bar{\theta}\right\rangle \right)^{2}L\left(i\right)}{\sum_{i}L\left(i\right)}}
\end{equation}
We give error bars on the results that correspond to two standard
deviation, and thus a returned value for $\bar{\theta}$ that is 
\begin{equation}
\bar{\theta}_{r}=\left\langle \bar{\theta}\right\rangle \pm2\sigma_{\left\langle \bar{\theta}\right\rangle }
\end{equation}

\subsection{Diffuse spectra\label{sub:Diffuse}}

When out of the specular spot, the radiation is controlled by the
complex transfer through the media (slab ice and bottom snow). The
experimental samples were made of pure water slab ice, without impurity.
We generated the look up table for every measurement geometry at very high spectral
resolution ($4.10^{-2}\,\mbox{nm}$) as required by the variability
of the optical constants of water ice, and then down sampled it at
the resolution of the instrument ($2\,\mbox{nm}$). We sampled the
17,085 combinations of two parameters for the 39 different geometries
: $p_{1}$ the thickness of the slab from $0\,\mbox{mm}$ to $20\,\mbox{mm}$
(noted $i=[1,201]$) every $0.1\,\mbox{mm}$ (noted $\mbox{d}p_{1}$),
and $p_{2}$ the grain-size of the granular substrate from $2\,\mbox{\ensuremath{\mu}m}$
to $25\,\mbox{\ensuremath{\mu}m}$ every $1\,\mbox{\ensuremath{\mu}m}$
and from $25\,\mbox{\ensuremath{\mu}m}$ to $1500\,\mbox{\ensuremath{\mu}m}$
every $25\,\mbox{\ensuremath{\mu}m}$ (noted $j=[1,85]$ and the corresponding
$\mbox{d}p_{2}(j)$). The parameter space is thus irregularly paved
with $\mbox{d}p(i,j)=\mbox{d}p_{1}.\mbox{d}p_{2}(j)$. 

For the inversion, we used the same method as previously described,
with a likelihood function $L$ that is written as in equation \ref{eq:L}.
Two different strategies were adopted. First, we inverted each spectra
independently. 39 geometries were sampled (described in section \ref{sub:Ice-slabs-measurements}),
and thus we conducted 39 inversions for each sample. This time $d_{sim}$
and $d_{mes}$ are thus respectively the simulated and measured spectra.
Then $d_{sim}$ and $d_{mes}$ are $n_{b}$-elements vectors, where
$n_{b}$ is the number of bands (61 in that study) and $\overline{\overline{C}}$
is a $n_{b}\times n_{b}$ matrix. As previously (see Section \ref{sub:Specular}),
we considered each wavelengths independently, thus generating a diagonal
matrix, containing the level of errors given by the technical data
of the instrument given by \cite{Brissaud2004}. The error is a percentage
of the measurement, and thus $\overline{\overline{C}}$ will be different
for every inversion.

Secondly, we inverted the BRDF as a whole, for each sample. For this
method, $d_{sim}$ and $d_{mes}$ are respectively the simulated and
measured BRDF, and thus are $n_{b}\times n_{geom}$-elements vectors
(2379 in that study), where $n_{b}$ is the number of bands (61 in
that study) and $n_{geom}$ is the number of geometries (39 in that
study) sampled. $\overline{\overline{C}}$
is a $\left(n_{b}\times n_{geom}\right)\times\left(n_{b}\times n_{geom}\right)$
diagonal matrix, containing the error on the data. We represent the
results the same way as previously, but there are two parameters to
inverse. For the sake of readability, we plot the normalized marginal
probability density function for each parameter. We present here the
general method for the inversion of $n_{p}=2$ parameters : the slab
thickness and the grain-size of the substrate. The PDF for the two
parameters $p$ is described by :

\begin{equation}
\mathcal{P}\left\{ p(i,j)\right\} =\frac{L\left(i,j\right)\mbox{d}p\left(i,j\right)}{\sum_{i}\sum_{j}L\left(i,j\right)\mbox{d}p\left(i,j\right)}
\end{equation}
For a given parameter $p_{1}$, the marginal PDF of the solution is
: 
\begin{equation}
\mathcal{P}\left\{ p_{1}(i)\right\} =\frac{L'\left(i\right)\,\mbox{d}p_{1}\left(i\right)}{\sum_{i}\sum_{j}L\left(i,j\right)\,\mbox{d}p\left(i,j\right)}
\end{equation}
with $L'\left(i\right)=\sum_{j}L(i,j)\mbox{d}p_{2}(j)$. The best match is the value $p_{1}(i)$ with the highest probability. The marginal
PDF can be described by the mean : 
\begin{equation}
\left\langle p_{1}\right\rangle =\frac{\sum_{i}p_{1}\left(i\right)L'\left(i\right)\,\mbox{d}p_{1}\left(i\right)}{\sum_{i}\sum_{j}L\left(i,j\right)\,\mbox{d}p\left(i,j\right)}
\end{equation}
and the associated standard deviation : 
\begin{equation}
\sigma_{\left\langle p_{1}\right\rangle }=\sqrt{\frac{\sum_{i}\left(p_{1}\left(i\right)-\left\langle p_{1}\right\rangle \right)^{2}L'\left(i\right)\,\mbox{d}p_{1}\left(i\right)}{\sum_{i}\sum_{j}L\left(i,j\right)\,\mbox{d}p\left(i,j\right)}}
\end{equation}
As for the roughness parameter, we give error bars on the results
that correspond to two standard deviation, and thus a returned value
for $p_{1}$ that is 
\begin{equation}
p_{1r}=\left\langle p_{1}\right\rangle \pm2\sigma_{\left\langle p_{1}\right\rangle }
\end{equation}

\begin{figure}[t]
\includegraphics[width=8.3cm]{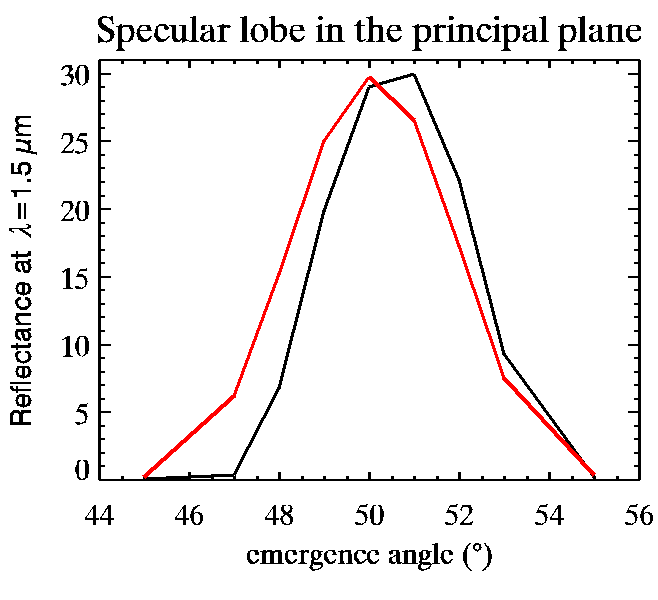}
\centering{(a)}
\includegraphics[width=8.3cm]{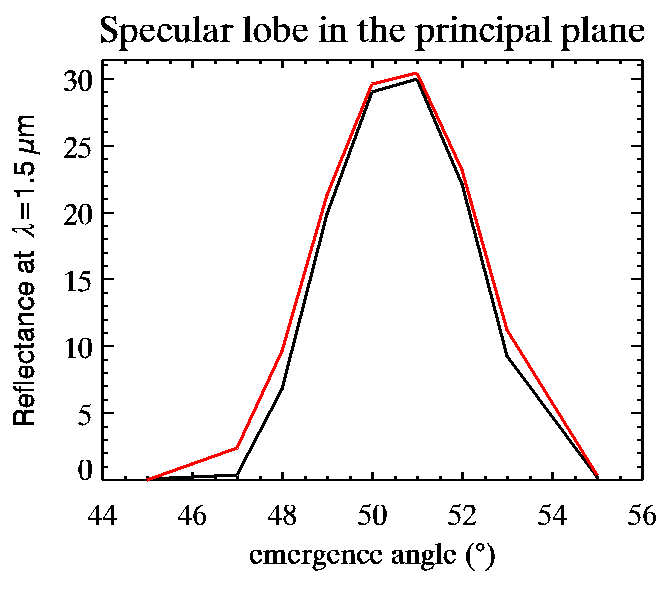}
\centering{(b)}
\caption{(a) Measured (black) and simulated (red) reflectance at $1.5\,\mbox{\ensuremath{\mu}m}$
in the principal plan for an incidence angle of $50\text{\textdegree}$.
The specular lobe measured is not centered at $50\text{\textdegree}$.
The sample may be slightly misadjusted resulting in a general drift
on the observation. (b) Measured (black) and simulated (red) reflectance
at $1.5\,\mbox{\ensuremath{\mu}m}$ in the principal plan for an incidence angle of $50\text{\textdegree}$.
We simulated a small miss-adjustment of the sample, resulting in a
shift of the observation of $0.5\text{\textdegree}$ in emergence
and $0.2\text{\textdegree}$ in azimuth. With this adjustments, the
model reproduce well the data. }
\label{fig:Measured-and-simulated1}
\label{fig:Measured-and-simulated2}
\end{figure}

\begin{figure}[t]
\includegraphics[width=8.3cm]{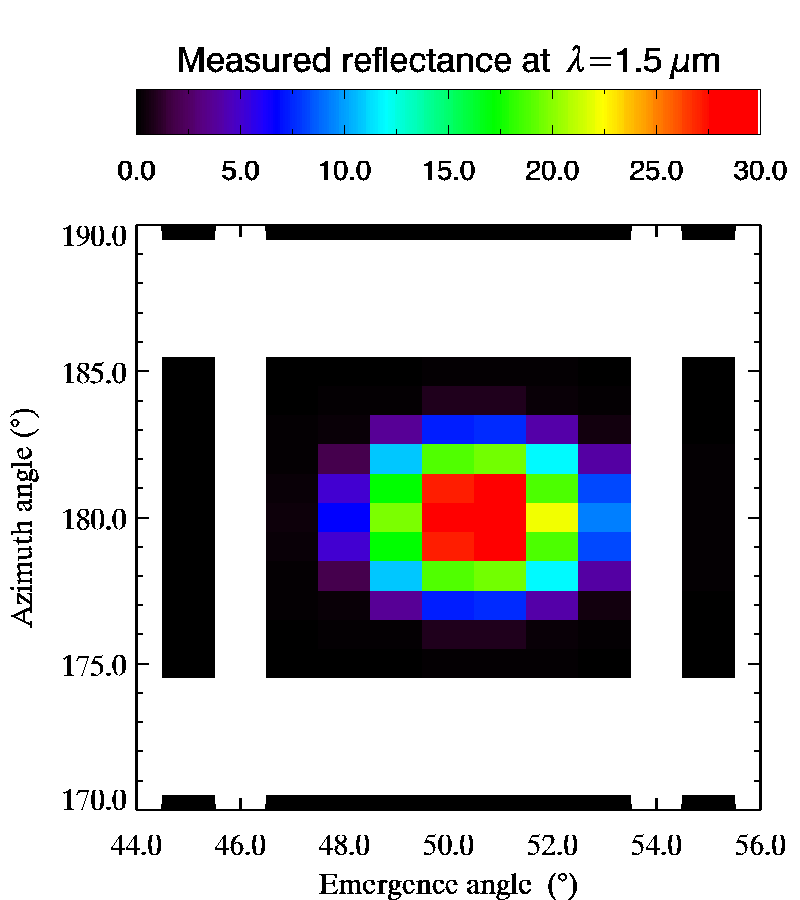}
\centering{(a)}
\includegraphics[width=8.3cm]{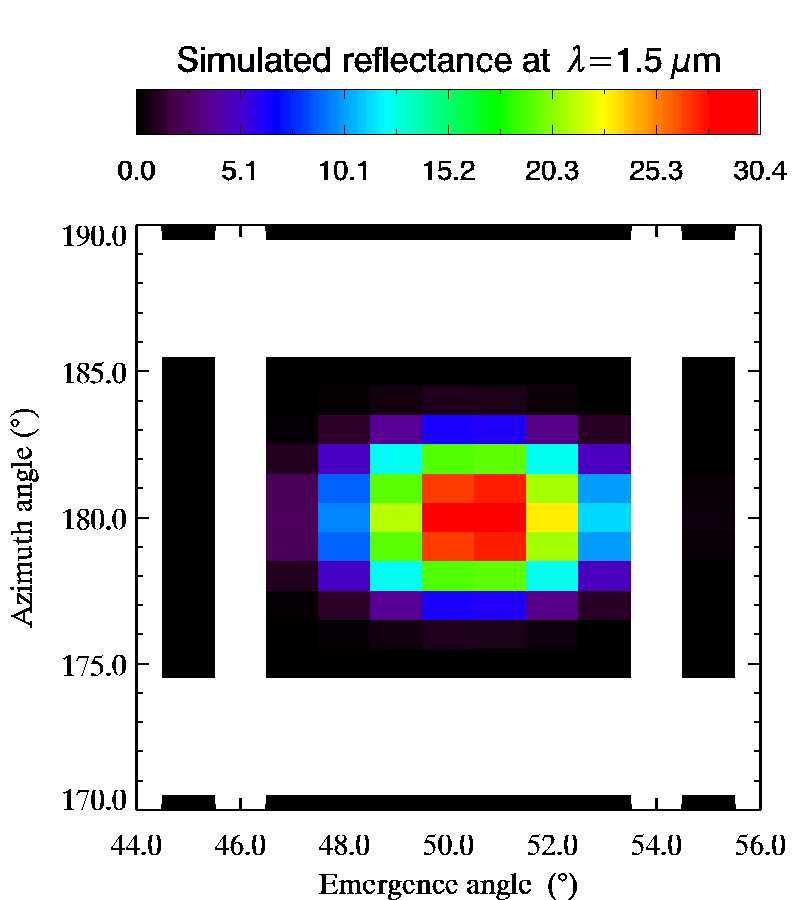}
\centering{(b)}
\caption{\label{fig:Specular2D}Measured and simulated reflectance around the
specular geometry at $1.5\,\mbox{\ensuremath{\mu}m}$ for an incidence angle of $50\text{\textdegree}$.
The simulation was computed assuming the determined shift of $0.5\text{\textdegree}$
in emergence and $0.2\text{\textdegree}$ in azimuth. The shape as
well as the intensity of the specular lobe are well reproduced. }
\label{fig:Specular2D}
\end{figure}

\begin{figure}[t]
\includegraphics[width=8.3cm]{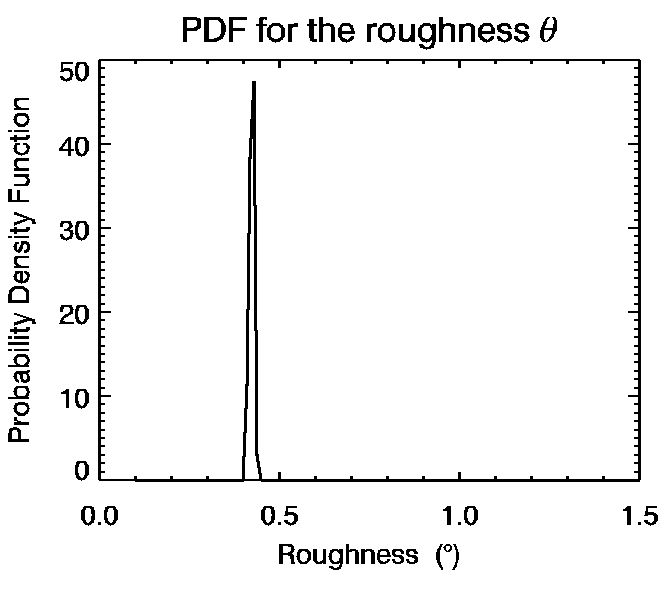}
\caption{\label{fig:Probability-density-function-theta}Probability density
function \emph{a posteriori} for the roughness parameter $\bar{\theta}$,
noted $\mathcal{P}\left\{ \bar{\theta}\right\} $. This function is
very sharp and thus the parameter $\bar{\theta}$ is well constrained.
The inverted value at $2\sigma$ is $\bar{\theta}=0.424\text{\textdegree}\pm0.046\text{\textdegree}$.
The best match is obtained for $\bar{\theta}=0.43\text{\textdegree}$.}
\label{fig:Probability-density-function-theta}
\end{figure}

\section{Results\label{sec:Results}}

\subsection{Specular lobe\label{sub:Specular-1}}

We performed the inversion taking into account 63 angels measurement,
but for the sake of readability, figure \ref{fig:Measured-and-simulated1} represents only the reflectance in the principle plane. The shapes and the intensities on Figure \ref{fig:Measured-and-simulated1}a are compatible, but measurement
and simulation are not centered at the same point. The simulation
is centered at the geometrical optics specular point (emergence $5\text{0\textdegree}$
and azimuth $180\text{\textdegree}$), whereas the measurement seems
centered around an emergence of $50.5\text{\textdegree}$. This could
be due to slightly mis-adjustment of the rotation axis of the sample
in the instrument. This kind of mis-adjustment are common, and can
easily result in a notable shift up to $1\text{\textdegree}$ of the
measure. We simulated different possible shifts in this range, and
found a best match represented on Figure \ref{fig:Measured-and-simulated2}b
for a shift of $0.5\text{\textdegree}$ in emergence, as it was suggested
by the first plot on figure \ref{fig:Measured-and-simulated1}a, and
$0.2\text{\textdegree}$ in azimuth. The measurements and the best
match are represented on figure \ref{fig:Specular2D}.
The shape as well as the magnitude of the specular lobe are very well
reproduced. Both lobes show a little asymmetry forward. This asymmetry
is not due to the sampling as it is present also when the simulation
is not shifted (see the red curve on figure \ref{fig:Measured-and-simulated1}).
It is due to an increase of the Fresnel reflexion coefficient when
the phase angle increase for this range of geometries. Figure \ref{fig:Probability-density-function-theta}
shows the PDF \emph{a posteriori }for the parameter $\bar{\theta}$.
The best match was obtained with $\bar{\theta}=0.43\text{\textdegree}$.
The inversion method gives a results close to gaussian shape at $\bar{\theta}=0.424\text{\textdegree}\pm0.046\text{\textdegree}$.
Unfortunately, we have no direct measurements of $\bar{\theta}$.
It would require a digital terrain model of the sample that is difficult
to obtain in icy samples. Still we find a low value, that is consistent
with the production in laboratory of slabs of columnar ice, that is
very flat, but still imperfects as described in the dataset. The average
slope is compatible with a long wavelength slope at the scale of the
sample, demonstrating that the micro-scale was not important in our
case. Indeed, for a sample that has a length $L$, an $1\sigma$ standard
deviation on the thickness $\Delta h$ can be attributed to a general
slope $\text{\ensuremath{\vartheta}}=\arctan\left(\frac{\Delta h}{L}\right)$
due to a small error in the parallelism of the two surfaces of the
slab. In the case of sample 3, $L=20\,\mbox{cm}$ and $\Delta h=1.35\,\mbox{mm}$
result in $\vartheta=0.39\text{\textdegree}$, which is compatible
with the roughness given by the inversion. We thus think that what
we see is an apparent roughness due to a small general slope on the
samples, and that the roughness at the surface is much lower than
this value. 

Moreover, the value retrieved by the inversion is very well constrained
as the probability density function is very sharp. This means that
we have an \emph{a posteriori }uncertainty on the result that is very
low. The quality of the reproduction of the specular spot by the model
suggest that the surface slope description is a robust description
despite its apparent simplicity. Especially, one single slope parameter
is enough to describe this surface. 

\begin{figure}[t]
\center{\includegraphics[width=6.3cm]{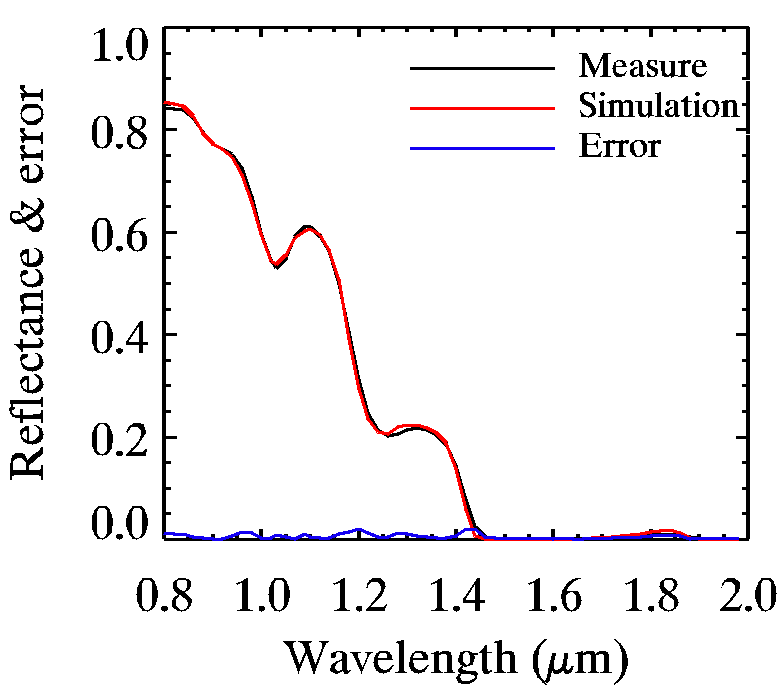}}
\center{(a) Sample 1. Thickness : $1.42\pm0.54mm$}
\center{\includegraphics[width=6.3cm]{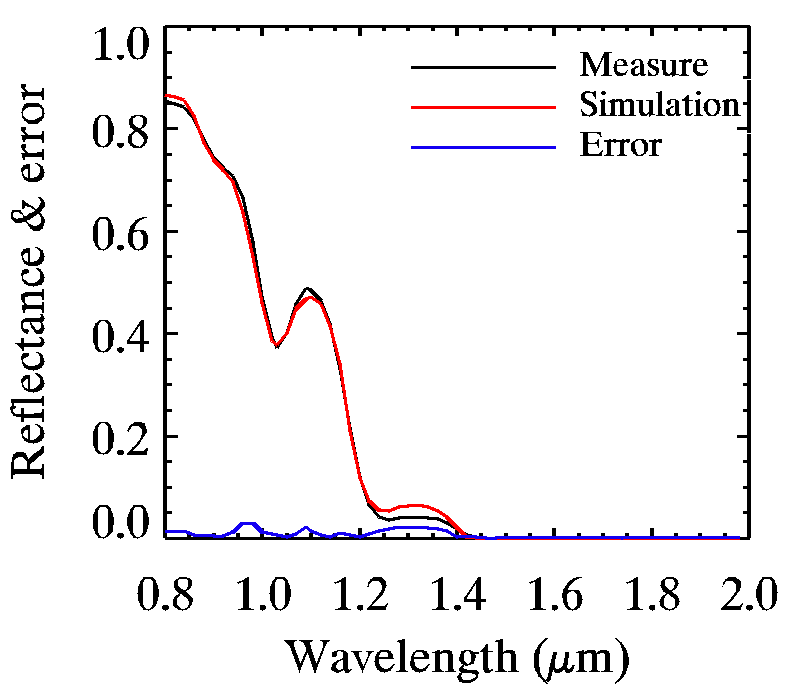}}
\center{(b) Sample 2. Thickness : $7.45\pm0.84mm$}
%\end{figure}
%\begin{figure}[t]
\center{\includegraphics[width=6.3cm]{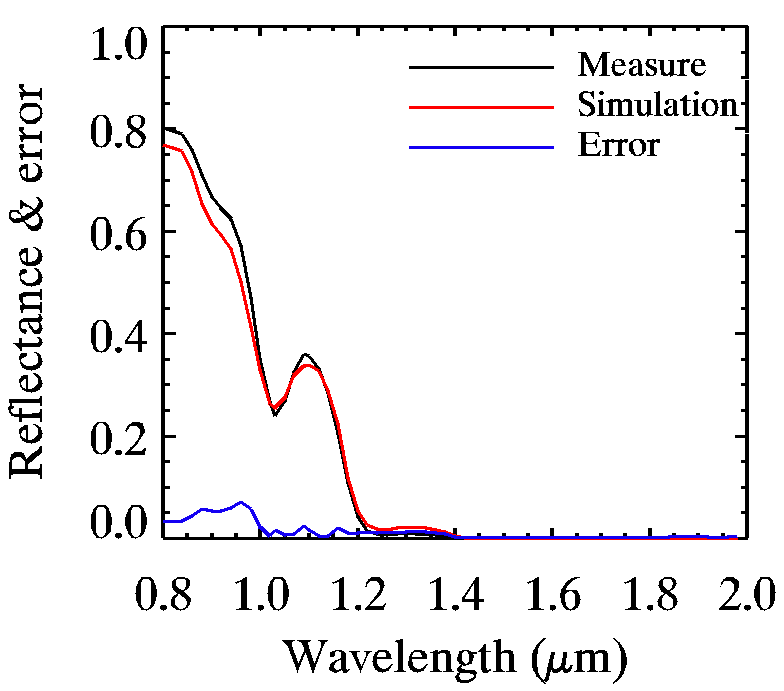}}
\center{(c) Sample 3. Thickness : $12.51\pm2.7mm$}
\caption{Measured and best match of simulated reflectance
spectra for the geometry of the best match for each sample : at incidence
$40\text{\textdegree}$, emergence $10\text{\textdegree}$ and
azimuth $140\text{\textdegree}$ for sample 1 (a), at incidence
$40\text{\textdegree}$, emergence $20\text{\textdegree}$ and
azimuth $45\text{\textdegree}$ for sample 2 (b), and at incidence
$60\text{\textdegree}$, emergence $0\text{\textdegree}$ for
sample 3 (c). The thicknesses indicated were measured before setting
the sample in the spectro-goniometer, and the errors are given at
$2\sigma$. The absolute differences are represented in blue on each
graph. The simulated spectra well reproduce the data within the range
of \emph{a priori }uncertainties. For the Sample 3 (c), the reflectance
in the $0.8\,\mbox{\ensuremath{\mu}m}-1.0\,\mbox{\ensuremath{\mu}m}$
range are not very well reproduced. The model cannot match the high
levels of the measurement. This could be explained by a change in
the experimental protocol, leading to the condensation of very fine
frost at the bottom of the slab layer.}
\label{fig:Fits}
\end{figure}

\begin{figure}[t]
\includegraphics[height=7.3cm]{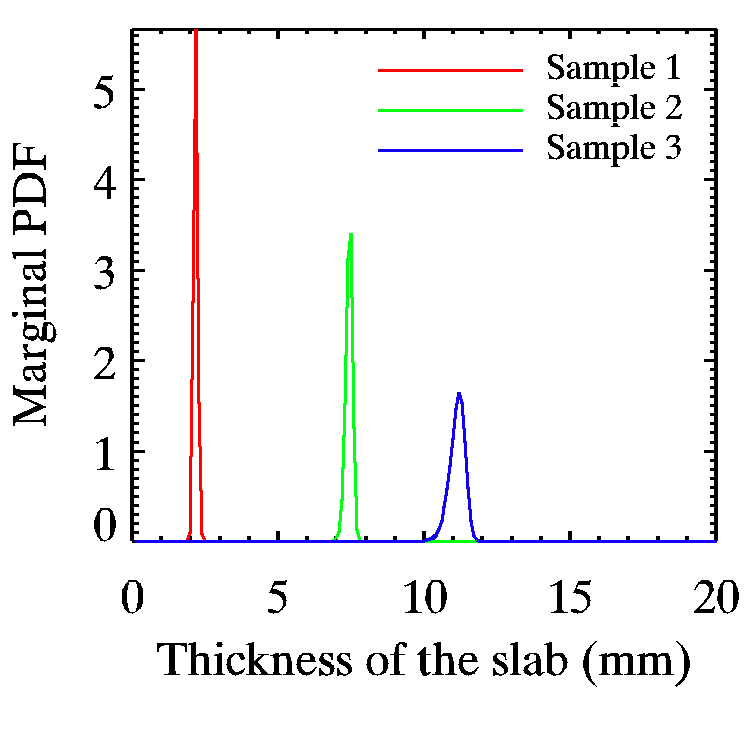}
\newline \centering{(a)}
\includegraphics[height=7.3cm]{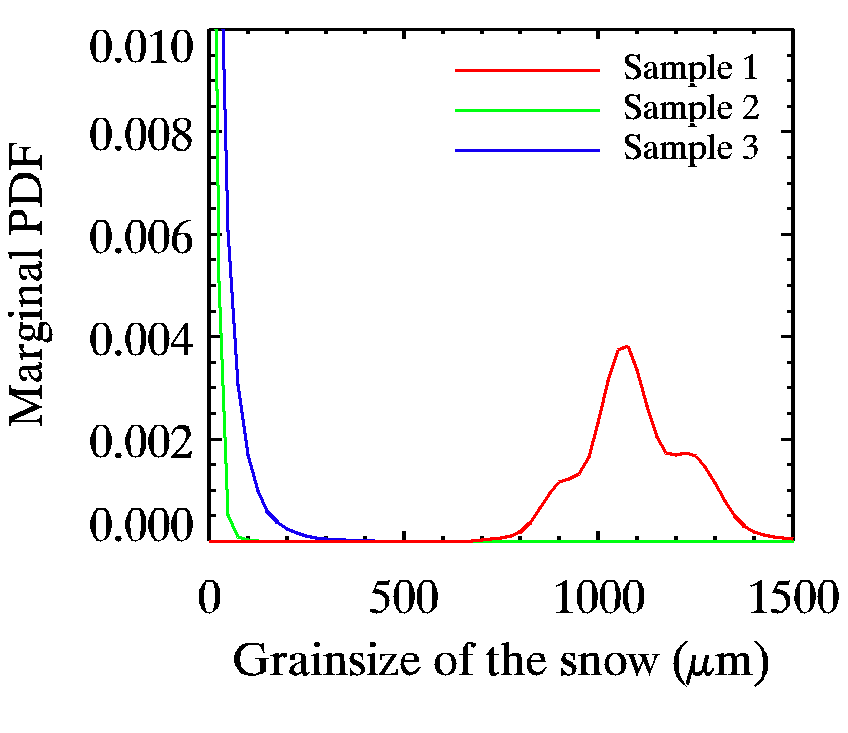}
\newline \centering{(b)}
\caption{Marginal probability density
functions\emph{ a posteriori} for (a) the thickness of the slab $\mathcal{P}\left\{ p_{1}(i)\right\} $
and (b) the grain-size of the snow substrate $\mathcal{P}\left\{ p_{2}(j)\right\} $
for the three samples, and for the geometries described in Figure
\ref{fig:Fits}. The functions are very sharp and very close to gaussian
for the thickness of the slab (a), but are broad for the grain-size
of the substrate (b). The thickness is well constrained by the inversion,
whereas the grain-size of the substrate cannot be determined with
high precision. }
\label{fig:Marginal-probability-density}
\end{figure}

\begin{figure}[t]
\includegraphics[width=8.3cm]{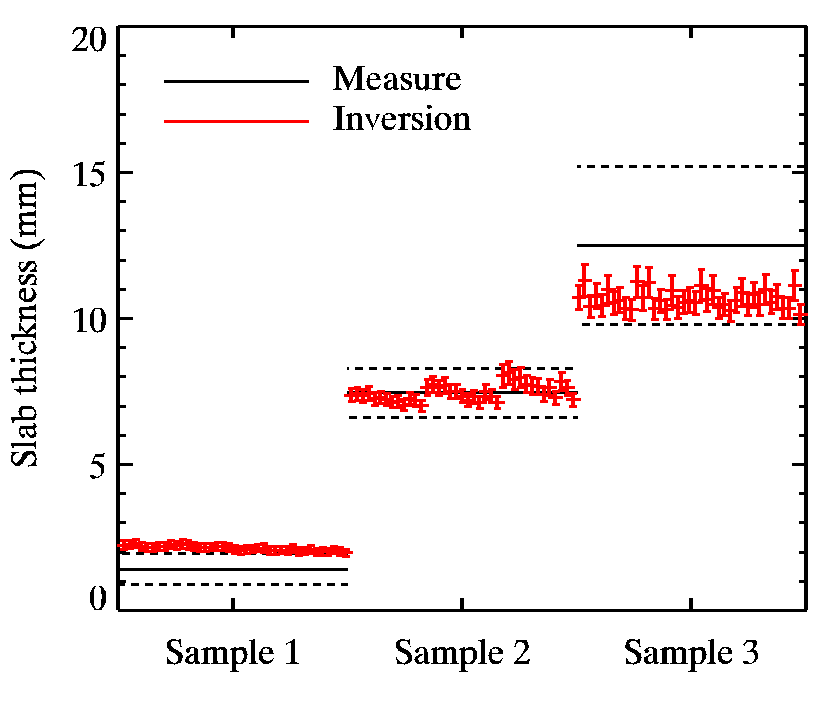}
\caption{Results of the inversion and measures with error
bars at $2\sigma$ for samples 1, 2 and 3, and for the 39 different
geometries of measurement. The thicknesses retrieved and measured
are compatible. The inversion points (in red) are sorted by incidence
(3 values), and each incidence is then sorted by azimuth (13 values
: 1 for emergence $0\text{\textdegree}$ and 6 for each $10\text{\textdegree\ and \ensuremath{20\text{\textdegree}}}$
emergences). The geometry seems to have an impact on the result for
sample 1 and 2. The thickness estimated seems to increase with incidence,
and decrease with azimuth. The geometrical effect seems to disappear
for big thicknesses.}
\label{fig:Results}
\end{figure}

\begin{figure}[t]
\center{\includegraphics[width=6.3cm]{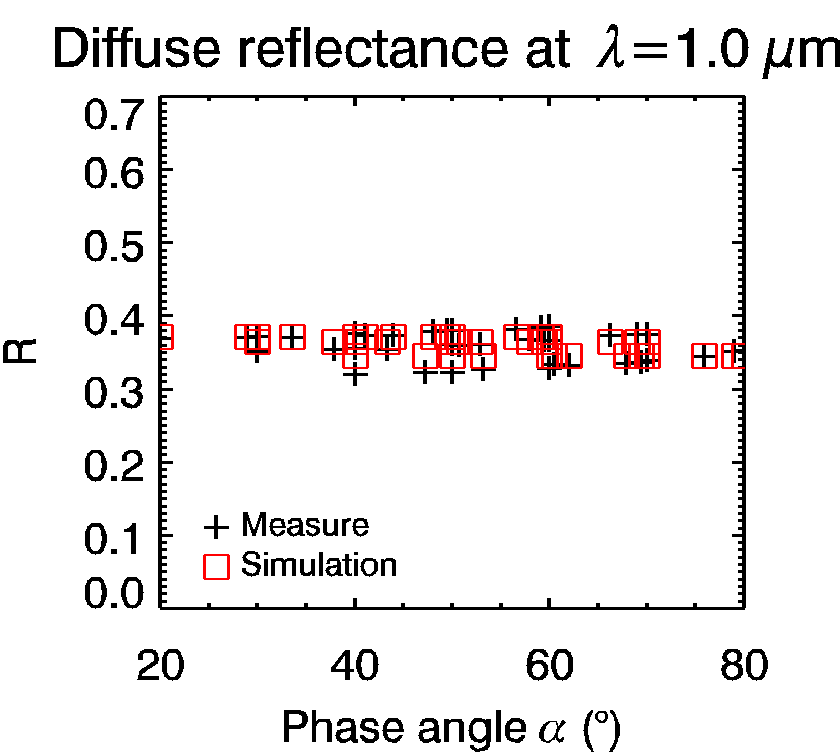}}
\center{(a) Sample 1. Thickness : $1.42\pm0.54mm$}
\center{\includegraphics[width=6.3cm]{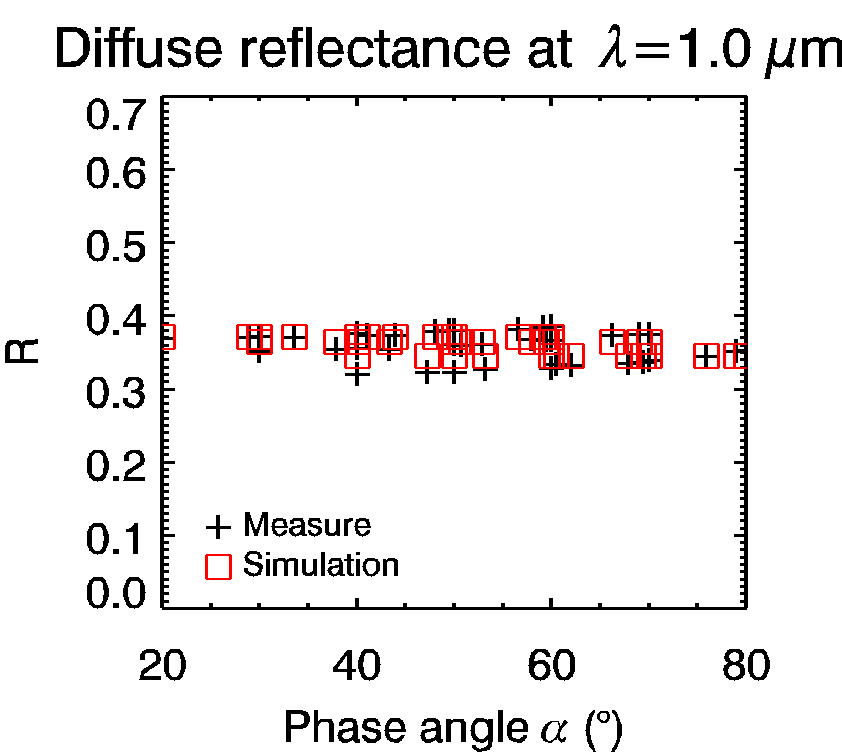}}
\center{(b) Sample 2. Thickness : $7.45\pm0.84mm$}
%\end{figure}
%\begin{figure}[t]
\center{\includegraphics[width=6.3cm]{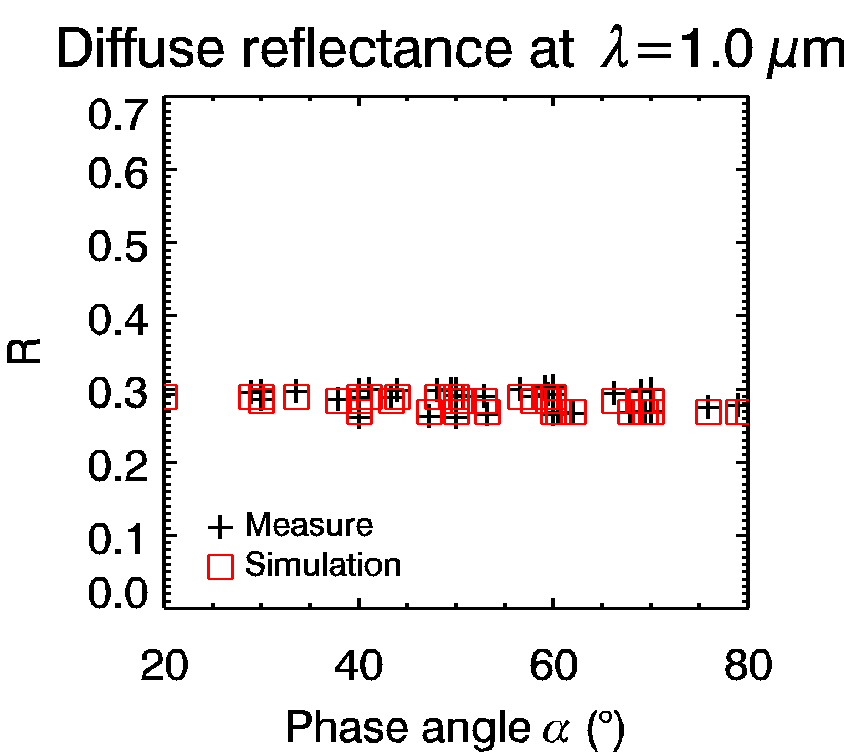}}
\center{(c) Sample 3. Thickness : $12.51\pm2.7mm$}
\caption{Measured and simulated reflectance factor
at $\lambda=1\,\mbox{\ensuremath{\mu}m}$ for (a) sample 1, (b) sample
2 and (c) sample 3. The simulation reproduces, if not perfectly, reasonably
the geometrical behavior of the surfaces. The quality of the geometrical
simulation seems to increase with the thickness of the slab. This
is consistent with the isotropization effect of a slab, that will
increase with the thickness. }
\label{fig:Meas-sim-phase}
\end{figure}

\begin{figure}[t]
\includegraphics[width=8.3cm]{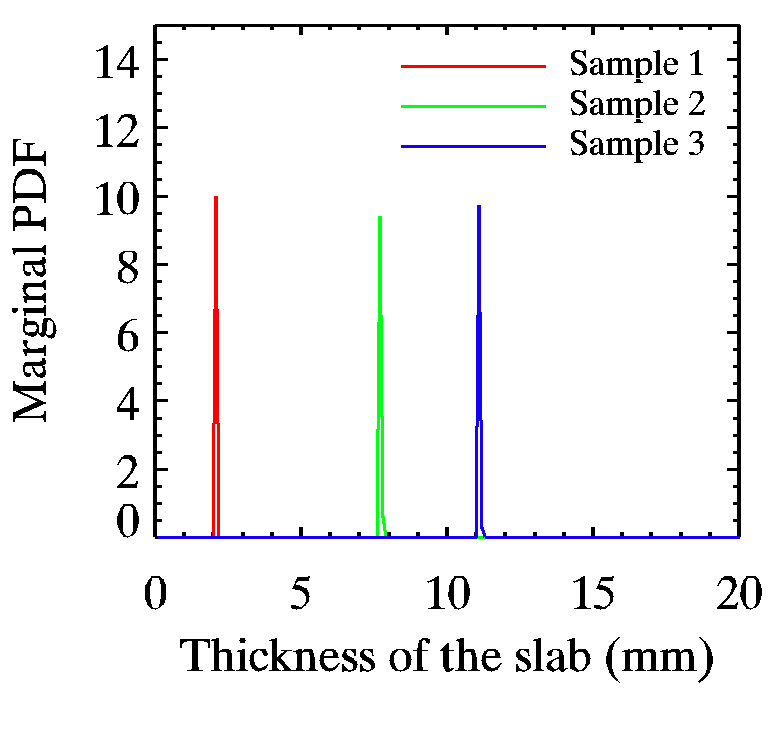}
\centering{(a)}
\includegraphics[width=8.3cm]{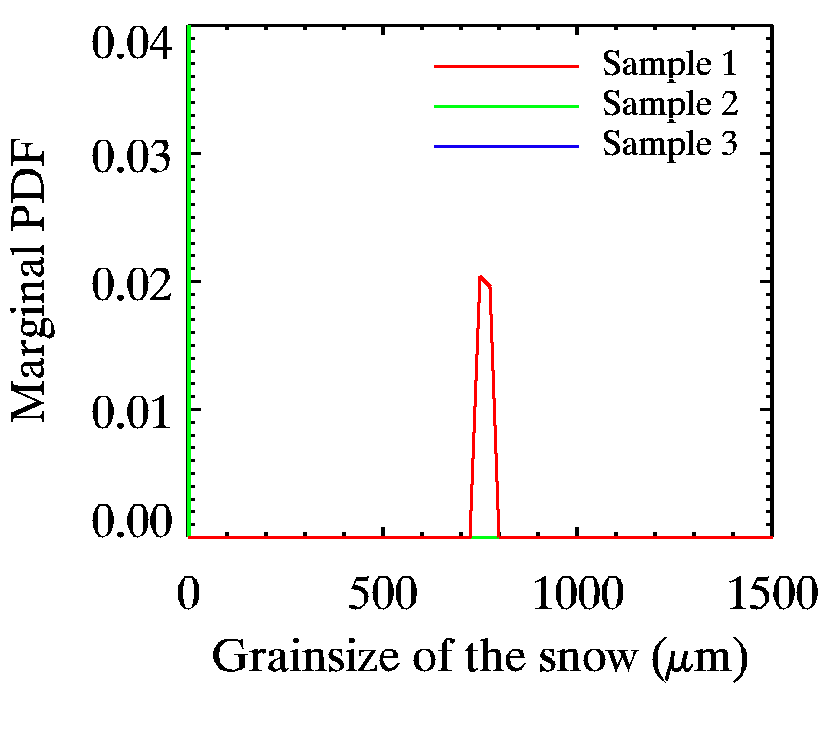}
\centering{(b)}
\caption{Marginal probability density
functions\emph{ a posteriori} for (a) the thickness of the slab $\mathcal{P}\left\{ p_{1}(i)\right\} $
and (b) the grain-size of the snow substrate $\mathcal{P}\left\{ p_{2}(j)\right\} $
for the three sample thickness. The functions are very sharp and very
close to gaussian for the thickness of the slab (a). The \emph{a posteriori}
uncertainties on the results are much smaller than the previous ones,
because the dataset is bigger and thus more constraining. Still, these
uncertainties are not fully reliable, as the model cannot reproduce
perfectly within the \emph{a priori} uncertainties the BRDF (see Figure
\ref{fig:Meas-sim-phase}). (b) The grainsize can be determined or
the sample 1, and is consistent with the results on inversions of
single spectra (see Figure \ref{fig:Marginal_g}). On the contrary,
they cannot be inverted for sample 2 and 3, as the returned probability
density function is close to a Dirac at the boundary of the definition
range. }
\label{fig:Marginal-probability-density-1}
\end{figure}

\subsection{Diffuse\label{sub:Diffuse-1}}

To reproduce diffuse reflectance we used the results obtained with
the specular measurement and assumed that the roughness of the samples
was not changing much between the experiments. The range of variations
in roughness should be negligible in the spectral analysis. We simulated
slabs over snow letting as free parameters the grain-size of the substrate
and the thickness of the slab. Figure \ref{fig:Fits} represent examples
of the best matches we obtained for the three measured samples at
various geometries. We also represented the mismatch between the best
fit and the observations. We find an agreement between the data and
the model that is acceptable. Nevertheless, there seems to be a decrease
of quality in the fits as the thickness increase. Figure \ref{fig:Marginal-probability-density}
shows on the same plot an example of the marginal PDF for the three
samples that are associated with the previous fits. Figure \ref{fig:Marginal-probability-density}
for the thickness of the slab (a) and the grain-size of the snow substrate (b). The thickness is well constrained as the marginal probability density functions \emph{a posteriori }are
relatively sharp and very close to gaussian. On the contrary, the
grain-size of the substrate seems to have a limited impact on the
result since it is little constrained. The marginal PDF for the grain-size
of the substrate are broad, and thus the\emph{ a posteriori} relative
uncertainty on the result is very high. Unfortunately, we have no
reliable measurement of the grain-size of the substrate, as it is
evolving during the time of the measurements. The general trend of
decreasing grainsize seems to be in agreement with eye's impression.

Figure \ref{fig:Results} show the measurements and the final result
of the inversion of the thickness for the three samples, and for 39
measurement geometries independently. The data and the model are compatible.
Still, the thickness of the sample 1 is slightly overestimated. This
may reveal a sensitivity limit of the model. The thickness of sample
3 seems underestimated. This could be partly due to the duration of
the measurement : the slab sublimates as the measure is being taken. Moreover, the specular measurements were performed on that sample increasing even more the duration of the experiment. The inversions points on Figure \ref{fig:Results} are sorted by increasing incidence, and for each incidence, by increasing azimuth. There seems to have an influence
of the geometry on the returned result : it is particularly clear
for sample 2. The estimated thickness tend to increase with incidence
and decrease with azimuth. This effect disappear for large thicknesses
(sample 3). Figure \ref{fig:Meas-sim-phase} show the measure and
the best match at the $\lambda=1.0\,\mbox{\ensuremath{\mu}m}$ wavelength,
when conducting the inversion on the whole BRDF dataset for each sample.
The relatively flat behavior of the radiation with the phase angle
is reasonably well reproduced. The quality of the geometrical match
increases with the thickness of the sample. This is consistent with
the fact that a thicker slab will permit a stronger isotropization
of the radiation. It is also consistent with the disappearance of
the geometrical dependence on the estimation for large thicknesses
noted on Figure \ref{fig:Results}. The values of thicknesses returned
by the inversion are displayed on Figure \ref{fig:Marginal-probability-density-1}a :
they are also compatible with the data, and the results are close
to the one given by independent inversions on each geometry (see Figures
 \ref{fig:Marginal-probability-density} and \ref{fig:Results}). The grainsizes returned,
even if compatible with the independent inversion results, are at
the boundary of the definition range of the parameter, for samples
2 and 3. This mean that the model cannot estimate this parameter correctly.
Indeed, as displayed on Figure \ref{fig:Marginal-probability-density-1}, the \emph{a
posteriori }marginal probability density functions for the samples
2 and 3 are very close to a Dirac at the lower limit of the domain.
This mean that the model inversion process cannot fit a value for
this parameter inside the definition domain that is fully satisfying.
This suggest an evolution of the conditions of the experiment between
the measure for the sample 1 and the others. The fact that the returned
value is at the lower boundary of the grain-size range suggest the
the actual grainsize of the snow is lower than this value. Unfortunately,
such grain-size would contradict the fundamental hypothesis of geometrical
optics assumed by the model. This results thus shall be interpreted
as grain-sizes smaller than the limit of detection. This kind of very
small grain-size could be produced during the experiments, by a small
temperature difference between the slabs and the natural snow, resulting
in the condensation of frost at the bottom of the slab layer.

\conclusions[Discussion and conclusion]
\label{sec:DiscussionConclusion}

The aim of this present work is to validate an approximate radiative
transfer model developed in \cite{Andrieu2015} using several assumptions. The
most debating one is the isotropization of the radiation when it reaches
the substrate. We first validated qualitatively this assumption with
snow and ice data. We then quantitatively tested and validated our
method using a pure slab ice with various thickness and snow as a
bottom condition. The thicknesses retrieved by the inversion are compatible
with the measurements for every geometry, demonstrating the robustness
of this method to retrieve the slab thickness from spectroscopy only.
The result given by the inversion of the whole dataset is also compatible
with the measurements. 

We also validate the angular response of such slab in the specular
lobe. Unfortunately, it was not possible to measure the micro-topography
in detail to compare with the retrieved data. Nevertheless, we found
a very good agreement between the simulation and the data. The average
slope is compatible with a long wavelength slope at the scale of the
sample, demonstrating that the micro-scale was not important in our
case. This is probably due to the sharp slicing method used. In future
work, an experimental validation of the specular lobe and roughness
should be addressed.

The large uncertainties on the grain size inversion demonstrate that
the bottom condition is less important than the slab for the radiation
field at first order, as expected. Even at a thickness of $1.4\,\mbox{mm}$,
since water ice is highly absorbent, the bottom layer is difficult
to sense. 

On figure \ref{fig:Fits}, there seems to be a decrease in the quality
of the fit when the slab thickness increase. We explain it by the
order in which the experiments were conducted. Indeed, the first measurement
was of the thinnest slab and the last on the thickest. In the meantime,
the snow substrate was sublimating. The errors in the fits could be
due to the increasing contamination in the substrate. The natural
snow cannot be perfectly pure : as it sublimates during the measurements,
the contribution of the contaminants become stronger and stronger.
These contaminant are not known, and not taken into account in the
model. A way to avoid this problem could be to set the slab ice on
top of a non volatile granular material, such as dry mineral sand,
which optical constants are known or can be determined. However this
would not solve another problem that is the re-condensation of water
into frost between the granular substrate and the slab. 

The comparison of the \emph{a posteriori }uncertainties on the thickness
of the slab and the grain-size of the snow substrate illustrate the
fact that those uncertainties depend both on the constraint brought
by the model itself and the uncertainty introduced in the measurement,
that only the bayesian approach can handle. The use of bayesian formalism,
is thus very powerful in comparison with traditional minimization
techniques.

We propose here an fast and innovative inversion method aiming at
massive inversion, for instance for remote sensing spectro-imaging
data, that enables to accurately estimate the uncertainties on the
model's parameters. As an example, the look-up table used for this
project were computed in $\sim150\,\mbox{s}$ for the roughness study
(1763 wavelengths sampled, 30933 spectra), and $\sim2.5\,\mbox{h}$
for the thickness and grainsize study (33,186 wavelengths sampled,
666,315 spectra). The inversion itself were performed in less than
one tenth of a second for specular lobe and independent spectral inversions,
and $2\,\mbox{s}$ for BRDF-as-a-whole inversions. Every calculation
was computed on one $4\,\mbox{GB}$ RAM Intel CPU. It has to be noted
that once the lookup table has been created, an unlimited number of
inversion can be conducted. The model is fast and the inversion is
highly parallel and thus adapted to the study of the compact ice covered
surfaces of the Solar system. For inversion over very large databases,
the code has been adapted to GPU paralellization. It is also possible
to increase de speed of the calculation of the look-up tables by doing
a multi-CPU computing.

%% REFERENCES

%% The reference list is compiled as follows:
%\bibliographystyle{plain}
%\bibliography{/Users/francoisandrieu/Documents/Biblio}
\bibliographystyle{copernicus}
\bibliography{Biblio}

\end{document}